\documentclass[12pt, draftclsnofoot, onecolumn]{IEEEtran}
\usepackage{graphicx}
\usepackage{subfigure}
\usepackage{cite}
\usepackage{amssymb, amsmath}
\usepackage{multirow}
\usepackage{epstopdf}

\usepackage{array}
\usepackage{breakurl}

\usepackage{color}
\usepackage{soul}
\soulregister\cite7
\soulregister\ref7

\begin{document}

\title{MEET: Mobility-Enhanced Edge inTelligence for Smart and Green 6G Networks }

\author{Yuxuan Sun,~\IEEEmembership{Member,~IEEE,}  ~Bowen Xie,
	~Sheng~Zhou,~\IEEEmembership{Member,~IEEE,}  ~Zhisheng~Niu,~\IEEEmembership{Fellow,~IEEE}
\thanks{Yuxuan Sun is with School of Electronic and Information Engineering, Beijing Jiaotong University, Beijing 100044, China, and was previously with Tsinghua University. Bowen Xie, Sheng~Zhou (\emph{Corresponding Author}) and Zhisheng~Niu are with Beijing National Research Center for Information Science and Technology, Department of Electronic Engineering, Tsinghua University, Beijing 100084, China.}
}

\maketitle

\begin{abstract}

Edge intelligence is an emerging paradigm for real-time training and inference at the wireless edge, thus enabling mission-critical applications. 
Accordingly, base stations (BSs) and edge servers (ESs) need to be densely deployed, leading to huge deployment and operation costs, in particular the energy costs.
In this article, we propose a new framework called \emph{Mobility-Enhanced Edge inTelligence (MEET)}, which exploits the sensing, communication, computing, and self-powering capabilities of intelligent connected vehicles for the smart and green 6G networks. Specifically, the operators can incorporate \emph{infrastructural vehicles} as movable BSs or ESs, and schedule them in a more flexible way to align with the communication and computation traffic fluctuations.
Meanwhile, the remaining compute resources of \emph{opportunistic vehicles} are exploited for edge training and inference, where mobility can further enhance edge intelligence by bringing more compute resources, communication opportunities, and diverse data. In this way, the deployment and operation costs are spread over the vastly available vehicles, so that the edge intelligence is realized cost-effectively and sustainably. Furthermore, these vehicles can be either powered by renewable energy to reduce carbon emissions, or charged more flexibly during off-peak hours to cut electricity bills. 
\end{abstract}

\section{Introduction}

6G networks are expected to support numerous mission-critical applications, such as autonomous driving, smart city, and industrial Internet of things. Artificial intelligence (AI)-based algorithms are widely involved in these applications, and stringent delay and reliability requirements need to be satisfied for communication and computation.

Integrating AI and edge computing technologies, edge intelligence is considered as an emerging paradigm to drive these applications \cite{Zhou2019edge}. User equipments (UEs) can offload their delay-sensitive and computation-intensive AI tasks to edge servers (ESs) for \emph{edge inference}, while UEs and ESs can also generate intelligence collaboratively from big data in a distributed and online manner via \emph{edge training} \cite{Sun2020edge}.
To enable the generation, dissemination and utilization of edge intelligence in real-time, base stations (BSs) and ESs need to be densely deployed. This will lead to huge deployment and operation costs, in particular the energy costs, which will bring heavy burdens to the operators.

Until 5G, operators have been deploying fixed-location BSs according to the peak traffic demand.
By the end of 2021, over 1.4 million 5G BSs have been installed in China. 
Although the energy consumed per bit data in 5G  is much lower than that of 4G, the power consumption of one 5G BS increases by 2 to 3 times. As a result, it is predicted that when 5G is fully deployed in China, the total power consumption of mobile networks will double, and the electricity bills for operators will be extremely high. If we further consider ESs, the total energy costs of the intelligent edge will increase significantly.  

Green communication and networking has been an important research topic during the last decade, and 
is drawing increasing attention nowadays \cite{Wang2022Green}. 
Existing solutions for 5G energy saving mainly include traffic-aware shutdown of sub-carriers, channels or whole BSs, and incorporating more renewable energy. However, the current idea is still to deploy sufficient fixed-location BSs to satisfy the peak traffic demand, and then seek energy saving opportunities during operation. Is this a desirable and sustainable solution? Should we still follow the same idea for the 6G deployment? 

While raising these questions, we also notice the rapid development of autonomous driving and vehicle-to-everything (V2X) communications. As more and more vehicles are equipped with powerful compute, communication and sensing capabilities, they can act as movable BSs and ESs to provide edge computing services \cite{Zhou2019exploiting}, or generate intelligence through distributed data collection and collaborative training. 


\begin{figure*}[!t]
	\centering
	\includegraphics[width=\linewidth]{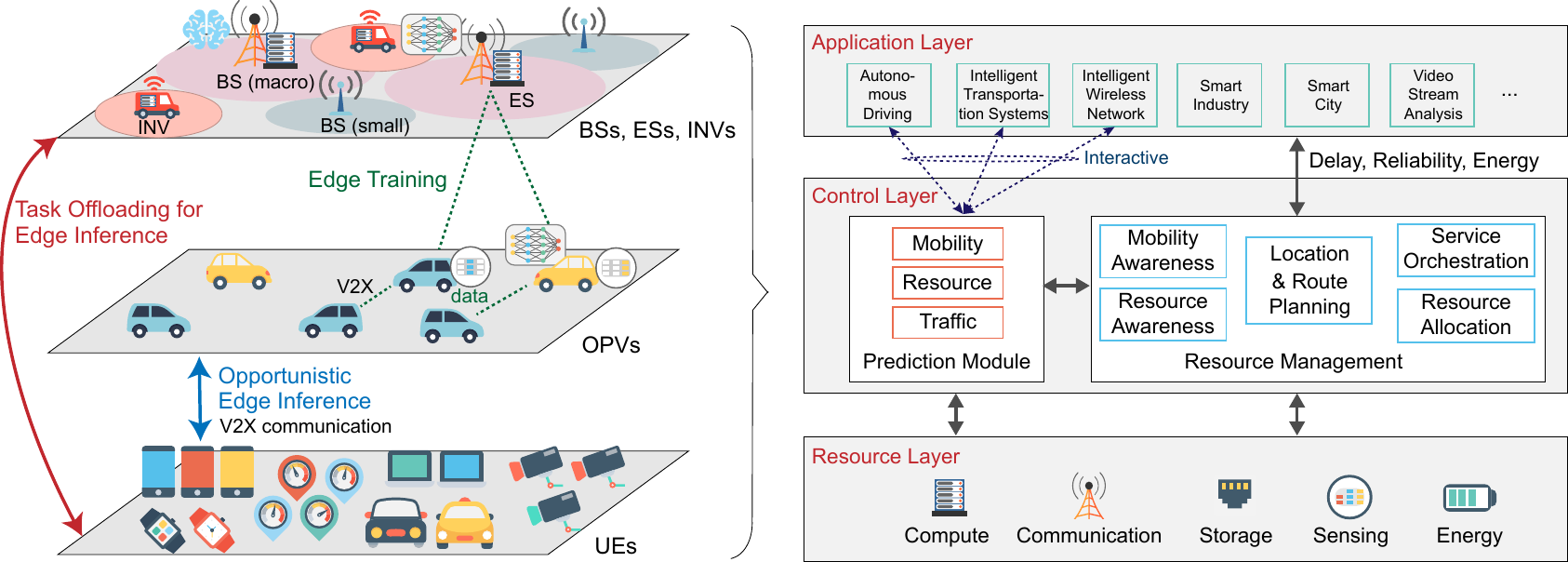}
	\caption{The MEET framework for edge training and inference.}
	\label{figsys}
\end{figure*}

In this article, we propose a new framework called \emph{Mobility-Enhanced Edge inTelligence (MEET)}, aiming to make 6G smarter and greener by incorporating vehicles. The operators can deploy dedicated vehicles as movable infrastructures, and dynamically manage their locations to \emph{meet} the fluctuated communication and computation traffic. Meanwhile, as the powerful compute platforms of autonomous vehicles do not need to run at the full load under normal road conditions, the remaining compute resources can be exploited to provide opportunistic computing services through \emph{task offloading}, or to train AI models with real-time sensing data. 
As vehicles will be powered by renewable energy or charged more flexibly, the edge intelligence can thus be realized cost-effectively and sustainably. 
While mobility is usually considered as a major cause of network performance degradation, we argue that it can be beneficial to the MEET system, and discuss the corresponding challenges and solutions. In specific, mobility may increase the probabilities to \emph{meet} more compute resources, communication opportunities, and diverse data, which can be exploited to enhance the performance of edge training and inference while saving the energy costs of the network.

\section{The MEET Framework} \label{sec_MEET}

The proposed MEET framework is shown in Fig. \ref{figsys}.
Two kinds of vehicles are considered: \emph{infrastructural vehicles} (INVs) and \emph{opportunistic vehicles} (OPVs). 
The INVs are the dedicated vehicles owned by the operators, and equipped with BS functionalities and edge computing capabilities. 
They act as movable BSs or ESs.
The OPVs are not deployed on purpose, but the vehicles with available compute resources and sensing capabilities, which can be exploited for opportunistic edge training and inference. According to the LTE or 5G V2X communication protocols, OPVs are considered as UEs who can communicate with BSs or INVs through the Uu interface, and with other UEs through the direct sidelinks using the PC5 interface \cite{Harounabadi2021V2X}.

As shown in Fig. \ref{figsys}, the logical structure consists of three layers: resource layer, control layer, and application layer. 
In the resource layer, the available compute, communication, sensing and storage resources, as well as the battery states of vehicles are abstracted and then imported to the resource management module in the control layer. 

In the control layer, there is a prediction module to forecast the mobility of vehicles, the communication and computation traffic fluctuations, and the availability of opportunistic resources using AI algorithms, such as long short-term memory network, recurrent neural network, graph neural network, etc. \cite{Liang2020Learning,Ma2022short}.
In specific, the mobility prediction includes two aspects. In the microscopic view, trajectory forecast aims to predict the location, velocity and steering of a vehicle in the next few seconds \cite{Liang2020Learning}. In the macroscopic view, the flow prediction aims to predict the traffic flow in a region \cite{Ma2022short}.
The prediction module interacts with the resource management module to dynamically optimize the deployment and plan the routes of INVs, as well as providing the awareness of opportunistic resources. Then, service orchestration and resource allocation are optimized for the training and inference tasks, while minimizing the energy costs of the network.
AI algorithms will be deeply integrated into the resource management module, in order to efficiently operate the network in the dynamic environment \cite{Huawei6G}.

As listed in the application layer, the MEET framework will support many AI-enabled mission-critical applications, such as autonomous driving, intelligent wireless network, smart city, etc., by providing timely edge intelligence. Vehicles can share their compute resources to execute inference tasks in real-time, and their sensing capabilities can also be exploited for data collection and timely edge training.
	

To measure the performance of the MEET system, we consider two kinds of performance metrics: the quality of service (QoS) of training and inference tasks, and the energy costs of the network.
Specifically, the QoS requirements mainly include delay and reliability. Inference tasks from mission-critical applications usually have stringent delay requirements. The total offloading delay of a task include both communication and computing parts.  The main goal is to minimize the task offloading delay, or to satisfy the reliability requirement by guaranteeing a high probability that tasks can be completed within a given delay deadline. 
Regarding the training tasks, our goal is to digest the fresh data in a timely manner by training accurate AI models with minimum delay.
Meanwhile, reducing the energy costs of the network is another key requirement. The MEET framework can disperse the electricity demands across time to cut electricity bills, and improve the utilization of renewable energy. 

For the implementation of the MEET system, V2X communication, mmWave communication, and integrated access and backhaul (IAB) technologies can be used for communication. Virtualization and containerization techniques are essential for the computing and storage aspects. Moreover, network function virtualization, software-defined networking, self-organized networking, and network slicing technologies can be exploited for the system management and control.

\subsection{Benefits from Mobility}

Conventionally, mobility is considered as a major cause of network performance degradations, since it brings more dynamics and uncertainties to the network topology, wireless channels, traffic distributions, etc. 
However, in the proposed MEET system, mobility may in turn make the network smarter and greener.
The potential benefits are as follows.


\noindent 1)  \emph{Reduction of deployment and operation costs}

The communication and computation traffic of the network fluctuates in both temporal and spatial domains, meaning that the network does not need to satisfy the peak traffic demands at all places simultaneously. Based on the monitoring and prediction of traffic demands, the locations and routes of the INVs are dynamically planned and adjusted. These INVs can confront the regional and momentary surges in service requirements. Traffic loads can also be offloaded to the OPVs. With higher mobility, BSs and UEs are more likely to meet more OPVs within a time period, and more capable vehicles may appear with higher probabilities. Accordingly, the total number of BSs and ESs to be deployed is reduced. As the energy consumption of an INV is lower than that of a BS (For example, Huawei eLTE Rapid system can be carried by the INV, whose power consumption is below 500W, while a 4G or 5G BS consumes at least 1kW), the total deployment and operation costs can be saved. 


Moreover, increasing number of vehicles will be powered by renewable energy in the future, which help reduce the carbon emissions of the network and save the energy costs. Compared with fixed-location BSs or ESs powered by renewable energy, it is more flexible to charge the movable INVs and OPVs.


\noindent 2) \emph{More communication and computing opportunities via sidelinks}

V2X communication capabilities using the sidelinks are being enhanced in the new releases of 3GPP \cite{Harounabadi2021V2X}, making it possible for an OPV to directly communicate with other OPVs or UEs for task offloading and collaborative training. OPVs may also act as relays to enhance the network coverage and improve the transmission rate.
With mobility, OPVs can meet more opportunities to communicate with BSs and UEs, and the transmission rates can be high due to shorter distances, better channel states, line-of-sight paths, etc. Accordingly, the communication delay and the transmit power are reduced. 
On the other hand, while mobility decreases the contact duration between OPVs and fixed-location BSs or UEs, the topology of vehicles may be relatively stationary if their velocities are close. Offloading tasks to the neighboring OPVs is thus a promising way, beneficial from higher data rates, less frequent handovers,  and lower transmit power.

\noindent 3) \emph{Advancements of data fusion}

Edge training tasks rely on distributed data. The non-independent and identically distributed (non-i.i.d.) data is a common challenge for most training tasks. Here, non-i.i.d. means that the distribution or feature of local data owned by an OPV is different from that of the global data, as well as that of other vehicles. Since vehicles sense data from different locations at different time epochs, their local data is naturally non-i.i.d. With mobility, the data sensed from different locations is carried to other places and disseminated to the whole network, and each OPV can also meet more OPVs from different places. This facilitates data fusion, and thus accelerates the convergence rate of training. Accordingly, the total energy consumption for training can also be reduced.

Some kinds of inference tasks, such as object detection and tracking, also require sensing data, where data fusion of multiple views may be beneficial. For example, cooperative perception is a promising solution to occlusions for autonomous driving, where neighboring vehicles share their sensing data to jointly percept objects. With mobility, an ego-vehicle is more likely to meet other OPVs who have clearer views on small or occluded objects, and thus the reliability of perception is improved. With reduced perception difficulty, it is also likely to use simpler detection algorithms to maintain the required reliability. The corresponding energy consumption can be reduced.

\subsection{Key Challenges}

To leverage the benefits of mobility, there are some key challenges to be addressed.

1) To minimize the deployment and operation costs while satisfying the QoS requirements, a key challenge is to optimize the densities of fixed-location BSs, ESs and INVs under different densities and mobility patterns of OPVs.  To solve this issue, a prerequisite is to analyze the fluctuations of communication and computation traffic based on big data and AI algorithms. 

2) Spectrum allocation and interference management is another key challenge for the movable BSs, which needs to be dynamically optimized. Joint prediction and optimization based on AI algorithms will be promising \cite{Huawei6G}, while beamforming and scheduling are efficient ways for interference management \cite{Wang2022Green}.

3) The limited bandwidth of the wireless backhaul links may become a bottleneck. Massive MIMO and mmWave band can be exploited. In fact, small cells also tend to use the wireless backhaul, such as IAB, to improve flexibility. 
The backhaul associations, bandwidth allocations and beam directions of vehicles should be jointly optimized whenever their locations change.

4) Regarding data fusion, a key challenge is to decide what to communicate given different channel conditions and contact durations. For example, we can optimize the splitting of a task to enable collaborative inference, and further compress the AI model or data.\cite{Sun2020edge}.

\section{The Deployment of Infrastructural Vehicles} \label{INV}
\begin{figure*}[!t]
	\centering
	\subfigure[Illustration of the deployment.]{\label{fig_INV_1}			
		\includegraphics[width=0.53\textwidth]{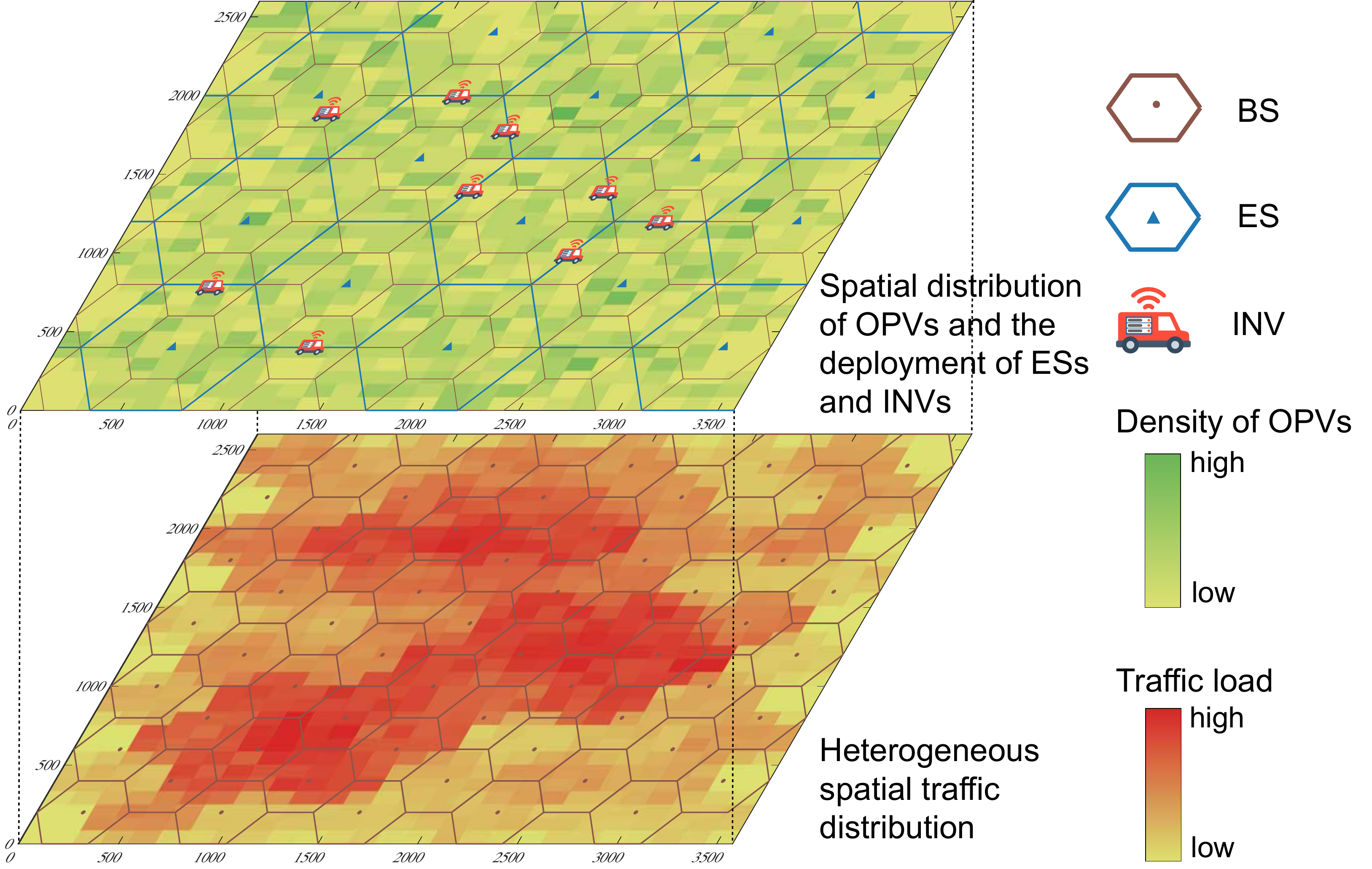}}
	\subfigure[Densities of ESs and INVs.]{\label{fig_INV_2}			
		\includegraphics[width=0.4\textwidth]{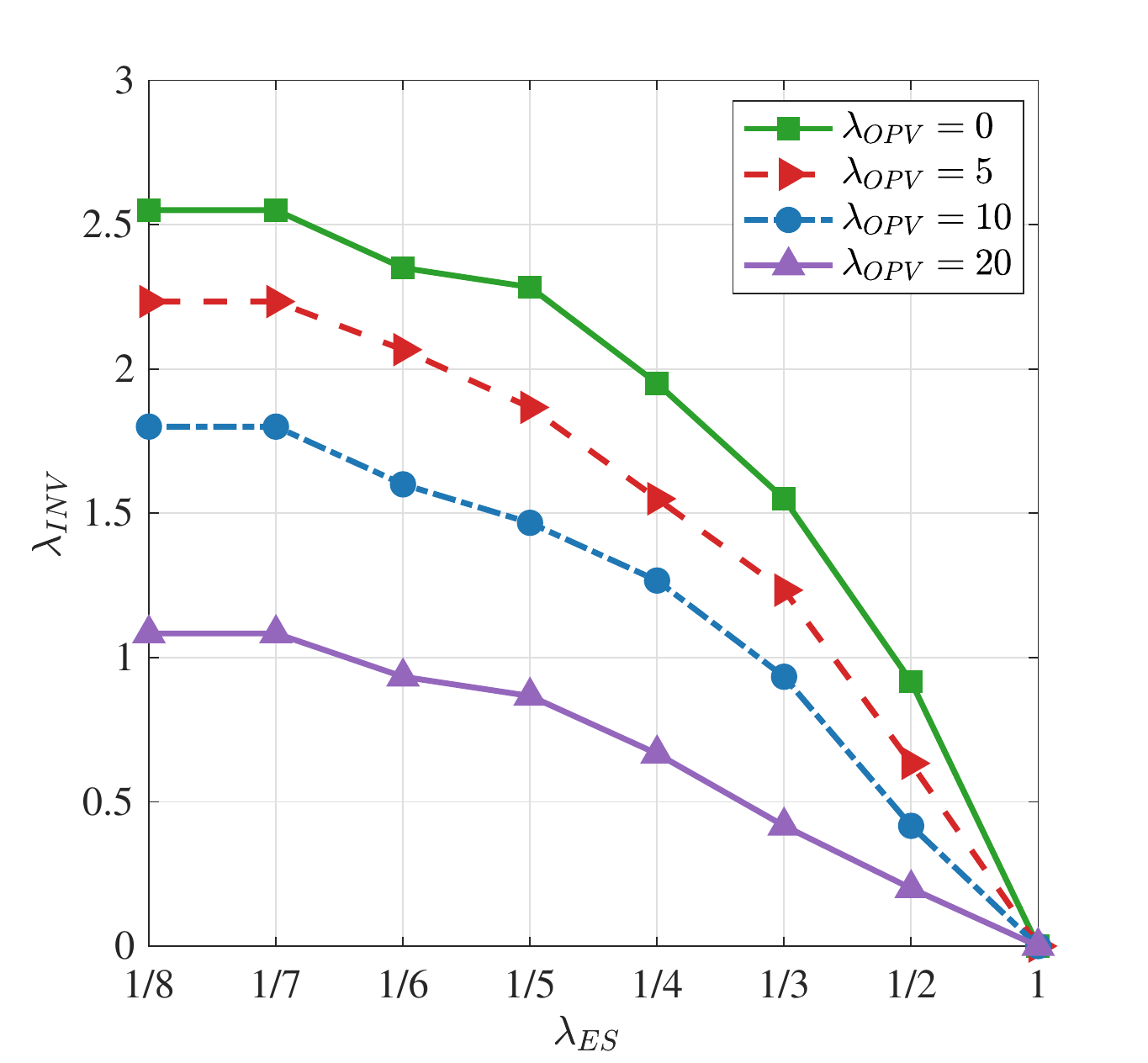}} 
	\caption{The joint deployment of INVs and ESs under different OPV densities.}
	\label{figexample}
\end{figure*}


In this section, we first show a numerical example of the cost-effective INV deployment, and then discuss some open problems.

Fig. \ref{figexample} provides an example of the joint deployment of ESs and INVs (acting as ESs) under different OPV densities. The BSs are deployed according to a hexagonal lattice, each covers a radius of 200m. The densities of ESs, INVs and OPVs, denoted by $\lambda_{\text{ES}}$, $\lambda_{\text{INV}}$, $\lambda_{\text{OPV}}$, respectively, are normalized by that of the BSs. The task arrival follows heterogeneous Poisson point process (PPP) in the spatial domain, whose normalized computation traffic intensity $\gamma_l$ is shown in Fig. \ref{fig_INV_1}. The OPVs follow homogeneous PPP. Assume that each ES, INV and OPV can serve 30, 10, and 1 task in each time slot, respectively. The goal is to jointly deploy ESs and INVs to guarantee that the blocking probability is below 0.1\%: $p_b(\lambda_{\text{ES}},\lambda_{\text{INV}}; \lambda_{\text{OPV}},\gamma_l)\leq 0.1\%$.
Given $\lambda_{\text{OPV}}$ and $\lambda_{\text{ES}}$, we can calculate $\lambda_{\text{INV}}$, as shown in Fig. \ref{fig_INV_2}. More OPVs and INVs can decrease the density of ESs. Based on the statistical information of OPVs and the deployment and operation cost ratio $\alpha$ of the INV and ES, we can then optimize the densities of INVs and ESs according to $\min \lambda_{\text{ES}}+\alpha\lambda_{\text{INV}}$.



The dynamic location optimization and route planning of INVs is still an open problem. As the predictions of traffic fluctuations cannot be perfectly accurate, it is crucial to design robust deployment schemes against unexpected traffic burstiness. Being aware of the energy state of each INV is also essential. 



\section{Exploiting Opportunistic Vehicles to Enhance Edge Intelligence} \label{sec_edge_computing}


The OPVs can be exploited to enhance edge intelligence via opportunistic inference and training. 
The key issue is how to seize the opportunities to communicate and compute to meet the QoS requirements of inference and training tasks. In this way, the network load can be offloaded to the OPVs greatly, and thus the energy costs of the network can be decreased.


\subsection{Opportunistic Task Offloading for Edge Inference} \label{subsec_computing}

Through task offloading, OPVs can provide various kinds of computing services for BSs and UEs.
As shown in Fig. \ref{figinfer}, we consider two types of task offloading schemes: infrastructure-to-vehicle (I2V) and vehicle-to-vehicle (V2V). In the I2V scheme, BSs offload tasks to the OPVs passing through. These tasks may be generated by the BSs themselves, or offloaded by the UEs. 
In the V2V scheme, UEs of driving systems, passengers or pedestrians directly offload their tasks to the neighboring OPVs via V2X communications. In this way, BSs do not need to transmit the data of tasks or collect the corresponding results. Compared with the I2V scheme,
both communication and computation workloads of BSs are reduced. 

To meet the stringent QoS requirements of inference tasks in a dynamic and distributed vehicular environment, it is important to design task partition and assignment algorithms. Meanwhile, since multiple OPVs may be available for a task, introducing \emph{redundancy} to task offloading may be beneficial via the \emph{diversity gain}. This can be achieved by task replication, which sends the copies of a task to multiple OPVs simultaneously, or coded computing, which brings tolerance to some OPVs with longer delay via coding techniques \cite{Zhou2019exploiting}. 
However, redundancy in turn increases the whole workloads of the network, and thus increases the communication delay and the queueing delay at each OPV. The optimal redundancy level should be optimized based on the amount of resources and service requests.

Deep neural networks (DNNs) are commonly involved in many AI-enabled applications to detect objects, make predictions, etc., where a tradeoff between delay and accuracy exists. A backbone architecture, such as ResNet or YOLOX \cite{Ge2021YOLOX}, can be configured with different depths. A deeper network improves the accuracy in general, but also increases the computational complexity. 
For example, YOLOX-X is over ten times more complex than YOLOX-S, but the average precision of object detection on COCO dataset is improved by 11.6\% \cite{Ge2021YOLOX}. 
Compression techniques such as pruning can further reduce the complexity of a DNN, and the input data can also be compressed to reduce communication workload \cite{Sun2020edge}. However, these compressions may also cause accuracy degradations. 
We need to optimize the DNN configurations and the compression schemes to balance the tradeoff between delay and accuracy, while fully exploiting the available resources of OPVs.


\subsection{Timely Edge Training with OPVs}

\begin{figure*}[!t]
	\centering
	\includegraphics[width=0.8\linewidth]{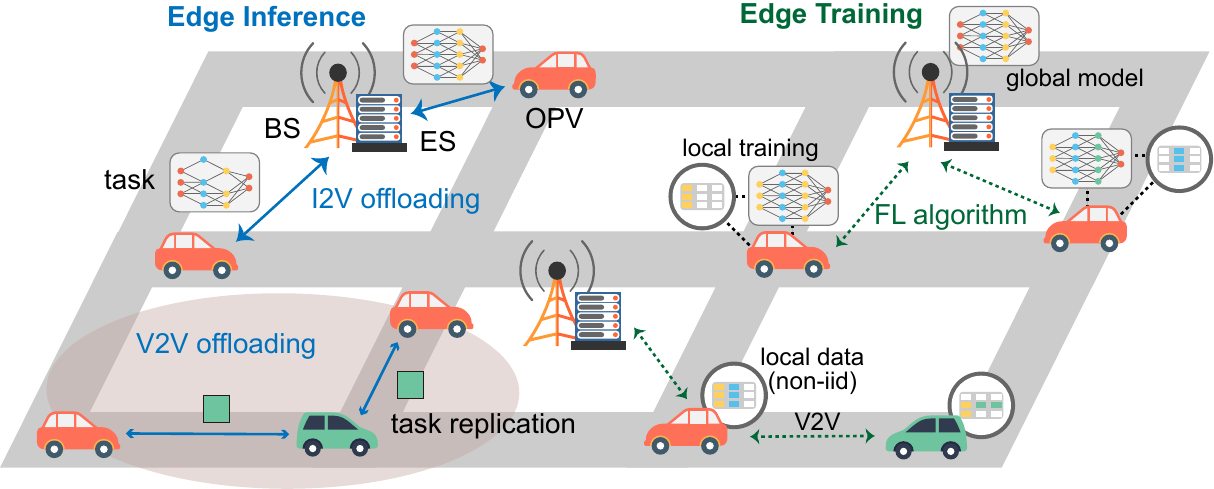}
	\caption{Exploiting OPVs for edge training and inference.}
	\label{figinfer}
\end{figure*}

In wireless networks, transportation systems, etc., the environments are usually time-varying. Accordingly, AI models need to be updated with fresh data in a timely manner. OPVs can be exploited to generate edge intelligence in a cost-effective way.

A prerequisite is to collect data samples online without manually labelling them. This is possible for many network management or prediction tasks. 
For example, in mmWave V2X communications, vision information can be employed by a DNN to predict beam strengths \cite{Mashhadi2021federated}. The corresponding dataset can be generated in an online manner, where OPVs carry out beam sweeping to search the optimal beams, and record the simultaneous positions, LiDAR and camera sensor data.
Another example is the motion forecasting task for autonomous driving, where an ego vehicle predicts the motions of surrounding vehicles and pedestrians \cite{Liang2020Learning}.  The real motions of these targets in the following time epochs provide the ground-truths naturally.

To train the AI models, we incorporate federated learning (FL) algorithm \cite{Zhou2019edge,Sun2020edge}, as shown in Fig. \ref{figinfer}.
A BS coordinates the training process by iterating the following steps: 1) When an OPV arrives, the BS transmits the current global AI model to it. 2) After receiving the model, the OPV trains it with its local data for multiple iterations, and then uploads the local model to the BS. 3) The BS collects the local models of OPVs, and averages them to update the global model once in a while. 
The key parameters that affect the timeliness of edge training are the number of iterations at each OPV, and the time interval of global model updates at the BS. 

Furthermore, multiple BSs can jointly train the model in a wider range using the hierarchical FL algorithm \cite{Feng2022Mobility}, where a central parameter server aggregates the models from BSs periodically. OPVs can also train the model in a decentralized and asynchronous manner without the help of BSs \cite{Zhou2019edge}.


\subsection{Key Research Issues and Possible Solutions}

\noindent 1) \emph{Diversity gain maximization for task offloading}

To fully exploit the OPVs and maximize the diversity gain, a key issue is to design redundancy-optimized task offloading algorithms under random arrivals of tasks and vehicles, and time-varying computing and communication capabilities. 

For the I2V offloading, we need to design online algorithms to decide which task in the queue should be offloaded upon the arrival of an OPV.
When the arrival rate of vehicles is relatively high, a task may be offloaded for multiple times to acquire the diversity gain. The goal is to maximize the ratio of tasks completed by the OPVs within the delay deadline. A Markov decision process based solution is proposed in \cite{Zhou2019exploiting} for homogeneous vehicles, while reinforcement learning based algorithms can be further designed for heterogeneous scenarios.

For the V2V offloading scheme, we would expect that offloading decisions can be made by each UE rather than the BS, since the topology of vehicles may be relatively stationary. Then, each UE can only acquire local state information about neighboring vehicles. However, optimizing the redundancy of tasks requires a global coordination of tasks and compute resources. Therefore, we have proposed a hierarchical optimization framework in \cite{Sun2021Replica}, where BSs optimize the redundancy level based on the statistical information of the global environment, and then each UE makes the instantaneous decision on its own. A case study based on \cite{Sun2021Replica} is shown in Section \ref{sec_case_study1}.

For both schemes, mobility-awareness is the key to the offloading performance. A lightweight mobility prediction method is proposed in \cite{Zaman2022LiMPO} for server selection and task offloading. A multi-hop task offloading policy is proposed in \cite{Liu2022mobility} for the V2V scenario based on the mobility patterns and computing capabilities. These algorithms can be further integrated when maximizing the diversity gain.

\noindent 2) \emph{Dynamic task orchestration for DNN tasks}

In terms of DNN-based inference tasks, we need to further optimize the compression ratios and the partition points, in order to satisfy both the delay and accuracy requirements. 
As discussed in Section \ref{subsec_computing} and \cite{Sun2020edge}, a lower compression ratio may increase the accuracy of an inference, but leads to higher delay. The optimal task orchestration needs to balance the trade-off between delay and accuracy. When redundancy is introduced, it is important to further coordinate the compression ratios across the replicas of each task by designing online algorithms.

\noindent 3) \emph{Scheduling policies for timely edge training}

To train an accurate AI model with minimum delay, we need to study how to coordinate the OPVs.
In FL, both aggregating more local models for a global update and carrying out more global updates improve the model accuracy, but  increase delay simultaneously. The convergence rate with respect to these two terms needs to be analyzed, based on which the interval of global model updates can be optimized. A case study is shown in Section \ref{sec_case_study2}.
For hierarchical FL, the central aggregation rule also needs to be designed based on the distribution and mobility of OPVs \cite{Feng2022Mobility}. 

To facilitate a smart yet green network, AI models should be trained \emph{on-demand}. The key is to monitor the accuracy of AI models and decide when to perform edge training. A possible way is to incorporate hypothesis testing based methods to detect the so-called \emph{concept drift}, which indicates the changes of underlying data distributions \cite{Lu2019learning}.

\section{Case Study I: Task Replication among Opportunistic Vehicles in MEET} \label{sec_case_study1}

In this case study, we demonstrate that OPVs can highly reduce the burden of the network through V2V offloading, by incorporating our previous work \cite{Sun2021Replica} into the MEET system. 
The goal is to maximize the ratio of inference tasks completed within the deadline, by optimizing the offloading redundancy and task assignment. The problem is solved in a hierarchical manner.
BSs collect the statistical information of the global environment, including the spatial distribution of OPVs, the task arrival rates and the service capabilities, and optimize the number of task replicas using the queueing theory.
Each UE receives the optimized redundancy periodically, and makes instantaneous offloading decisions in a decentralized manner. With only local information, it is difficult to know which neighboring OPVs can provide more reliable services. We thus incorporate \emph{combinatorial multi-armed  bandit} for online learning, and adapt it  to the dynamic topology \cite{Sun2021Replica}. 

The trajectories of vehicles are generated by Simulation for Urban MObility (SUMO, https://www.eclipse.org/sumo/). The maximum speed of each vehicle is 20\text{m/s}. The arrival of tasks follows Poisson process, and the deadline of each task is 0.2s. The density ratio of UEs over OPVs is 0.25, and the service delay at each OPV follows exponential distribution with rate 10 tasks per second. 

\begin{figure}[!t]
	\centering
	\includegraphics[width=0.6\linewidth]{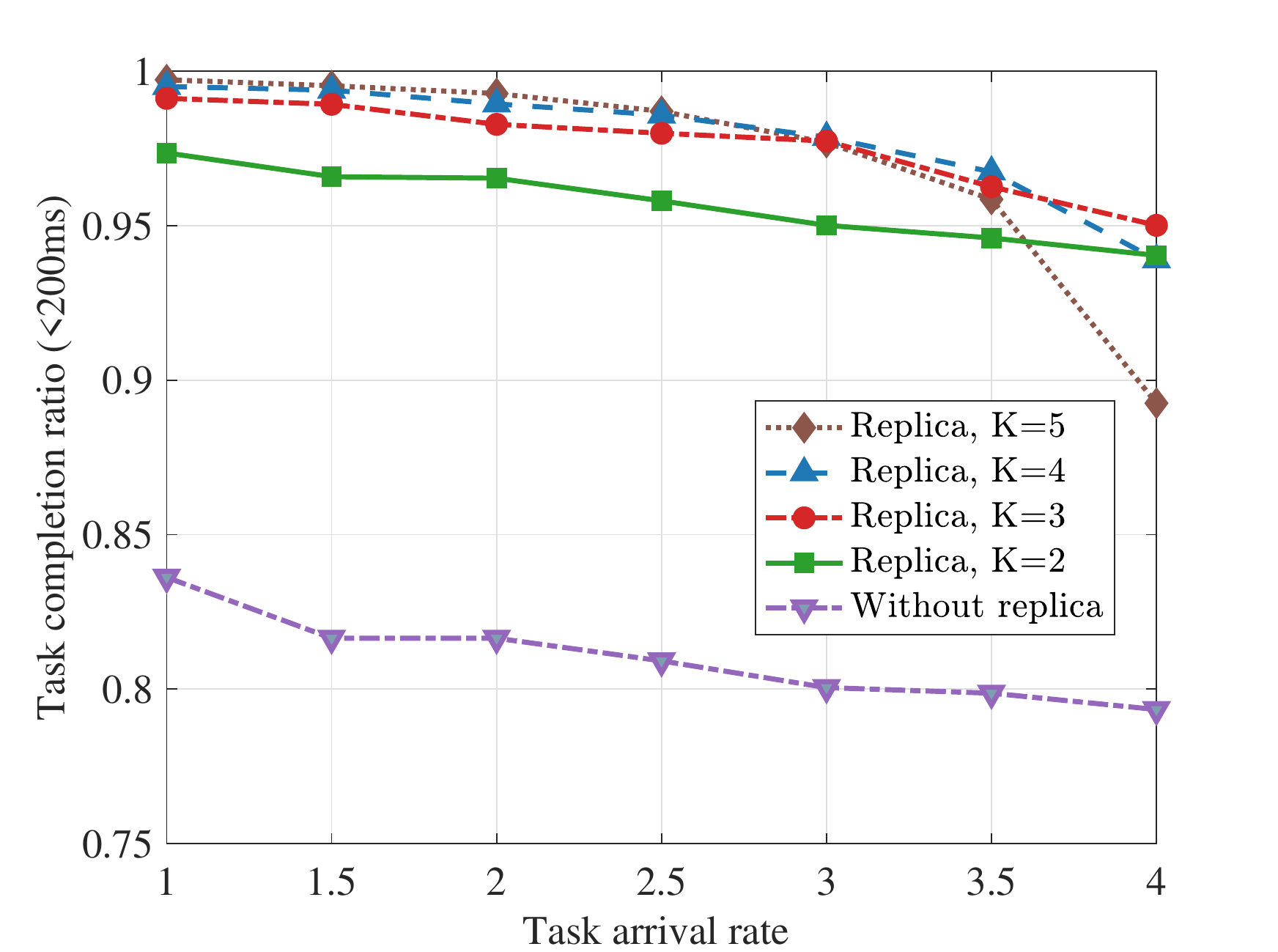}
	\caption{The ratio of inference tasks completed by the OPVs.}
	\label{case_study_replica}
\end{figure}

Under different number of replicas and task arrival rates, the task completion ratio within the deadline is shown in Fig. \ref{case_study_replica}. We find that redundancy is not the higher the better. With an optimized number of task replica, the proposed V2V offloading scheme can serve over 95\% of tasks. Compared to the case without replica, the task completion ratio is increased by around 15\%, which validates the diversity gain.

\section{Case Study II: Mobility-Enhanced Edge Training in MEET} \label{sec_case_study2}

\begin{figure}[!t]
	\centering
	\vspace{-3mm}
	\includegraphics[width=0.6\linewidth]{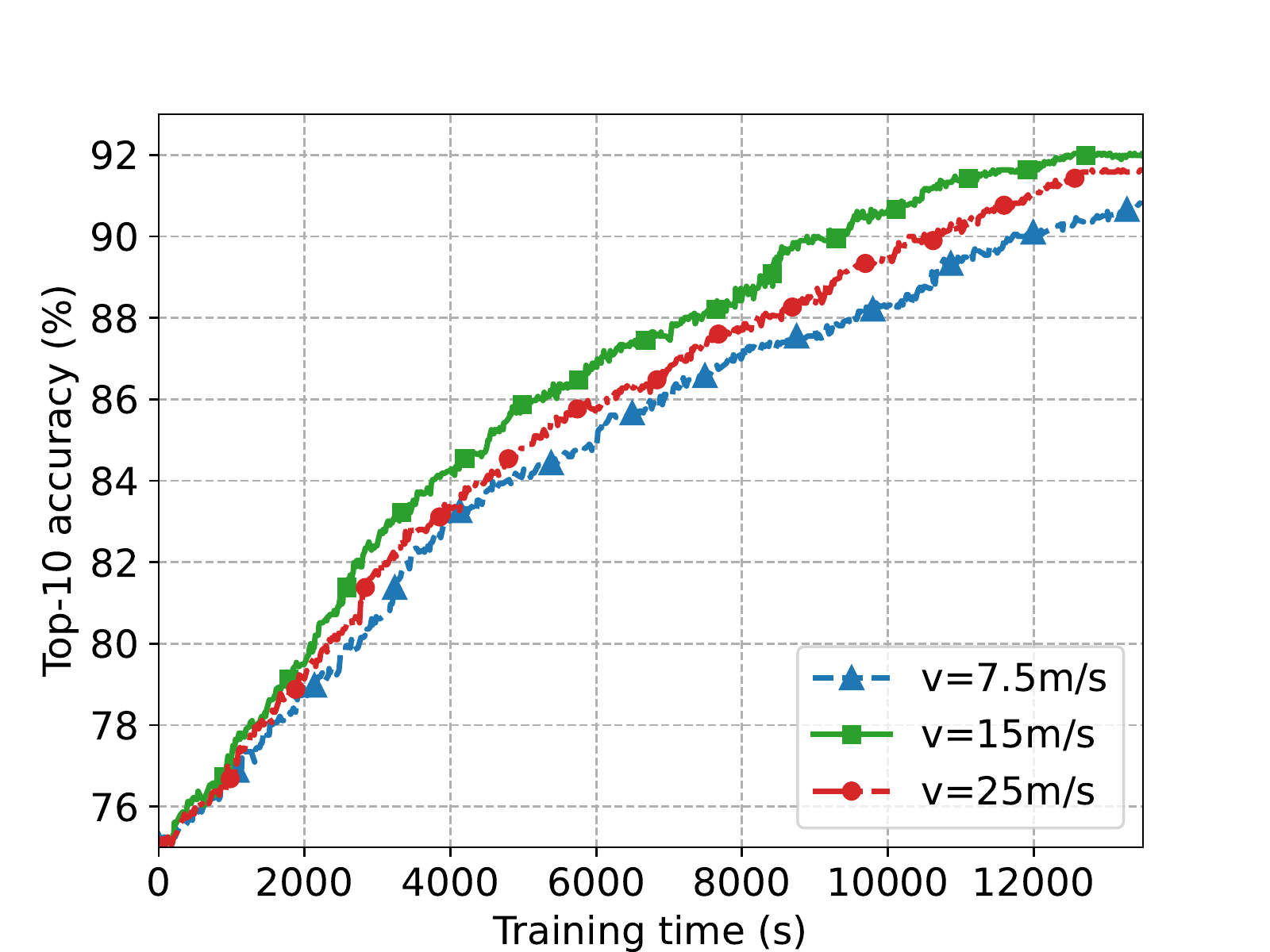}
	\caption{The top-10 accuracy achieved over time under different velocities.}
	\label{case_study_fl}
\end{figure}

In this case study, we show that OPVs can generated edge intelligence timely.
We consider an AI-enabled mmWave beam selection DNN trained by the OPVs in a federated manner.
A BS covers a road segment of 400m. Under its coverage, the arrival of vehicles follows Poisson process with rate $\lambda$. According to the traffic theory, $\lambda=\rho_{\text{max}}v\left(1-\frac{v}{v_{\text{max}}}\right)$, where $\rho_{\text{max}}=200$ vehicles per km is the maximum vehicle density, $v_{\text{max}}=35\text{m/s}$ is the maximum velocity allowed, $v$ is the average velocity, and then $\lambda$ can be calculated. Among these vehicles, we assume that 5\% OPVs involve in this training task (others may not have data or compute resources).

We adopt the same DNN architecture as \cite{Mashhadi2021federated} to infer the optimal beam using the LIDAR data of vehicles, and use the Raymobtime dataset (https://www.lasse.ufpa.br/raymobtime/) for training. 
As edge training does not learn from scratch in general, we use a pre-trained model with an initial top-10 accuracy of 75\%, meaning that the probability that the 10 beam directions predicted by the DNN contains the optimal one is 0.75.
The total training time is 13000s.
The BS broadcasts the model parameters to all the OPVs whenever a global update occurs. Each OPV carries out local training for 20 iterations, each with batch size 128, and then uploads the model to the BS. The total communication plus local training delay follows shifted exponential distribution with rate parameter $10 s^{-1}$ and shift parameter 0.1s. At the BS, the global model is updated when a local model is received. The learning rate is 0.1. 

The top-10 accuracy achieved over time under different velocities $v$ is shown in Fig. \ref{case_study_fl}. Different velocities affect the arrival rate of OPVs and the duration of coverage, and thus the convergence rate of training. Among the three chosen velocities, $v=15$m/s converges the fastest and achieves a top-10 accuracy of over 92\%, which is higher than around 90\% in \cite{Mashhadi2021federated}. Results demonstrate that OPVs can be exploited for edge training, and mobility can be beneficial.

\section{Conclusions}
We have proposed a MEET framework for smart and green 6G networks by exploiting the communication, computing, sensing, and self-powering capabilities of vehicles as well as their mobility. INVs are incorporated  into the wireless network to improve the flexibility and cost-effectivity of BS and ES deployments, while OPVs are exploited for opportunistic edge training and inference.
Through two case studies, we have shown that OPVs can help to reduce the workloads of ESs and BSs significantly, and generate the edge intelligence timely.
As future directions, security and privacy guarantees are obligatory, while designing incentive mechanisms to boost the participation of OPVs is another important direction.

\end{document}